\documentclass[11pt]{article}
\usepackage[dvips]{graphicx}
\usepackage{pdproc}

  \makeatletter 
  \def\@cite#1{[#1]} 
  \def\gtrsim{\mathrel{\rlap{\lower4pt\hbox{\hskip1pt$\sim$}}
    \raise1pt\hbox{$>$}}}         
  \makeatother    
\begin{document}

\renewcommand{\thefootnote}{\alph{footnote}}

\title{
 The Supersymmetric Fat Higgs
}

\author{ RONI HARNIK}

\address{ 
Physics Department, UC Berkeley, Berkeley, CA 94720, USA\\
Theoretical Physics Group, LBNL, UC Berkeley, Berkeley, CA 94720, USA
\\ {\rm E-mail: roni@socrates.berkeley.edu}}

\abstract{
Supersymmetric models have traditionally been assumed to be perturbative up
to high scales due to the requirement of calculable unification. In this note
I review the recently proposed `Fat Higgs' model which relaxes the
requirement of perturbativity. In this framework, an NMSSM-like trilinear
coupling becomes strong at some intermediate scale. The NMSSM Higgses are
meson composites of an asymptotically-free gauge theory. This allows us to
raise the mass of the Higgs, thus alleviating the MSSM of its fine tuning
problem. Despite the strong coupling at an intermediate scale, the UV
completion allows us to maintain gauge coupling unification.  }

\normalsize\baselineskip=15pt

\section{ Motivation - The Little Hierarchy Problem(s) }

The post-LEP era has forced models of electroweak symmetry breaking (EWSB) to
confront data.
For example, in models with a low cutoff one is expected to write all of the
operators allowed by symmetries, suppressed by the appropriate power of the
cutoff, $\Lambda$, in the spirit of EFT. Assuming these operators have
coefficients of $O(1)$, one can place a lower bound on the cutoff from EW
measurements $\Lambda \gtrsim 10 \mathrm{TeV}$~\cite{Barbieri:2000gf}.
However, since it is natural to tie the EW to the dynamical $\Lambda$, one
then needs to explain the little hierarchy between them. This is the `Little
Hierarchy Problem' or `LEP Paradox'. 

In supersymmetric models, where physics is weakly coupled at a TeV, one
naively does not run into this problem because there is no low cutoff at
which we must introduce all of the operators allowed by symmetry. However,
a little hierarchy problem emerges once the MSSM is required to fulfill
the LEP bound for the mass of the Higgs, $m_h>115$~GeV. In the MSSM the
Higgs' quartic coupling is determined by SUSY and is given by the $D$-term
potential. This places an upper-bound of $m_Z$ on the tree level Higgs mass.
One can raise this value by radiative corrections from top-stop loops, but
this requires the stop to be quite heavy, $m_{\tilde t}> 500 $~GeV. However,
top-stop loops are also responsible for EWSB in the MSSM, thus the stop mass
is involved in setting the EW scale. The need to address the little hierarchy
between $m_{\tilde t}$ and the EW scale is the `Supersymmetric Little
Hierarchy problem' \footnote{For a more detailed discussion see
e.g.~\cite{fathiggs}.}.
Within the MSSM it may only be solved by fine tuning other parameters in the
Higgs potential.

The tree level bound on the Higgs mass may be raised by going beyond the
MSSM. In the NMSSM an additional singlet is added along with a superpotential
\begin{equation}
 W = \lambda N H_u H_d - \frac{k}{3} N^3,
\end{equation}
which gives an additional quartic coupling to the Higgs, 
$\lambda^2|H_u H_d|^2$. This contributes to raise the Higgs mass to 
$m_h^2 \sim \lambda^2 v^2 + O(m_Z^2)$ which can easily be above 115 GeV.

However, we cannot raise the mass arbitrarily by increasing $\lambda$. The
reason is that the effective $\lambda(\mu)$ grows in the UV and eventually
hits a Landau pole at a scale $\Lambda$. Above this scale we lose control of
the theory and even the relevant degrees of freedom are unknown. Encountering
this scale below the GUT scale would be a disaster for unification. Demanding
that the Landau pole be above $M_{GUT}$ has yielded an upper-bound on
$\lambda(v)$ and thus on the Higgs mass. In the NMSSM this gives $m_h<150$
GeV.

However, in principle, nothing can prevent us from giving up unification and
bringing $\Lambda$ down to increase the mass of the Higgs.  This model is now
is an EFT below a cutoff which we imagine as some dynamical scale of
compositeness.  As we have seen, this model solves the supersymmetric version
of the little hierarchy problem. Since we are now dealing with a low-cutoff
model we may wonder if the ``regular'' little hierarchy problem was
re-introduced i.e.  explaining the little hierarchy between the cutoff and
the EW scale.  However this problem is trivially solved since the little
hierarchy is stabilized by supersymmetry. 

It is interesting to draw an analogy between the models of Little
Higgs~\cite{little} and the strongly coupled version of the NMSSM peresented
above.  Both models are EFTs with a low cutoff suplemented by additional
symmetries that protect the Higgs mass. In the former the symmetry is a
global symmetry that is collectively broken and in the later it is
supersymmetry. The main advantage of the supersymmetric model is that due to
exact results in strongly coupled supersymmetric gauge theories, it is much
easier to UV complete. 

\section{UV Completion -- The Fat Higgs}

In~\cite{fathiggs} we UV completed a cousin theory of the NMSSM given by the
superpotential
\begin{equation}
W= \lambda N(H_u H_d-v_0^2).
\label{fat}
\end{equation}
This superpotential is similar to the NMSSM in that it raises the tree level
Higgs mass above the LEP bound for a sufficiently large $\lambda(v)$ but then
requires a UV completion above the Landau pole of $\lambda$.

In this brief note, I will only present the essential components of our model
and advertise some of its features.
A more complete model and its analysis is presented in~\cite{fathiggs}.
The UV dynamics of our model consists of a new
strongly coupled gauge group, $SU(2)_H$. The we introduce six doublets,
$T^{1\ldots 6}$, under $SU(2)_H$ which corresponds to $N_f=3$. The $SU(2)_H$
is asymptotically free and thus becomes strong at some scale,
$\Lambda_H$. The $T$'s are also carry electroweak charge. The charge assignments under
$SU(2)_H\times SU(2)_L\times U(1)_Y$ is
\begin{equation}
(T^1, T^2) \equiv T = (\mathbf{2}, \mathbf{2},0), \qquad
(T^3, T^4) = (\mathbf{2}, \mathbf{1},\pm \frac{1}{\, 2}), \qquad
(T^5, T^6) = (\mathbf{2}, \mathbf{1},0).
\end{equation}
We also write a tree-level superpotential
\begin{equation}
\label{Wtree}
  W = - m T^5 T^6 +W_{decouple}
\end{equation}
where the meaning of $W_{decouple}$ will become clear presently. Below the
scale $\Lambda_H$ this theory confines and low energy degrees of freedom are
described by the anti-symmetric meson matrix $M_{ij}=T_i T_j$.  Let us
relabel part of the meson matrix as follows
\begin{equation}
  N = M_{56}, \,
  \left(
    \begin{array}{c}
      H_u^+ \\ H_u^0
    \end{array} \right)
  = \left(
    \begin{array}{c}
      M_{13} \\ M_{23}
    \end{array} \right), \,
  \left(
    \begin{array}{c}
      H_d^0 \\ H_d^-
    \end{array} \right)
  = \left(
    \begin{array}{c}
      M_{14} \\ M_{24}
    \end{array} \right),
\end{equation}
noting that the mesons labeled as $H_u$ and $H_d$ indeed transform as the
Higgs fields of the MSSM, and $N$ is a singlet, as needed in Eq.~(\ref{fat}).
The superpotential $W_{decouple}$ in Eq.~(\ref{Wtree}) involves additional
new fields that marry all of the other meson fields in $M_{ij}$. This can be
done naturally by enforcing an additional $Z_3$ symmetry~\cite{fathiggs}.
The theory below $\Lambda_H$ is thus an effective theory of $H_u$, $H_d$ and
$N$ alone.

This theory generates a dynamical superpotential 
\begin{equation}
W_{dyn}=\frac{\mathrm{Pf M}}{\Lambda^3}\supset
\frac{1}{\Lambda^3}NH_uH_d 
\end{equation}
Once $H_{u,d}$ and $N$ are canonically normalized to dimension one fields this
will produce the renormalizable trilinear term in Eq.~(\ref{fat}). The mass
for the third flavor of Eq.~(\ref{Wtree}) becomes a linear term for the
singlet $N$. Putting these two contributions together gives a superpotential
of the form of Eq.~(\ref{fat}). This superpotential breaks EW symmetry even
in the SUSY limit, which releases the stop from its role in EWSB in the MSSM.
The scale $v_0$ at which EWSB occurs in the SUSY limit may be estimated in
NDA as $v_0^2=\frac{m\Lambda_H}{(4\pi)^2}$ .

\begin{figure}[tc]
\centering \includegraphics[width=.43\columnwidth]{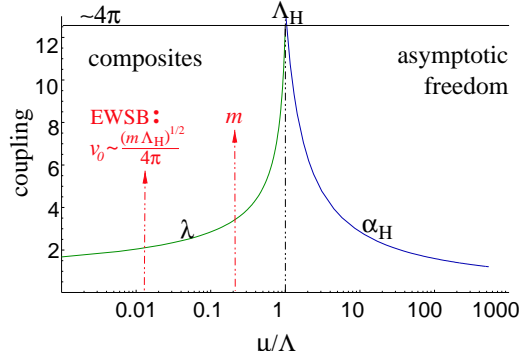}
\caption{The renormalization of the couplings in the model presented.  When
$4\pi v_0\ll \Lambda_{H}$ the mesons condense at weak coupling and the theory
is calculable.}
\label{running}
\end{figure}

The dynamics in this model are heuristically summarized in
figure~\ref{running}. At high energy the theory is a weekly couple gauge
theory with the $T$'s as the relevant degrees of freedom. As we run down
the theory becomes strong at $\Lambda_H$ at which point we substitute the
degrees of freedom to the meson fields and the coupling $\lambda$ for the
gauge coupling. Below $\Lambda_H$ the coupling $\lambda$ renormalizes quickly
to weak coupling. If the scale $m$ is somewhat below $\Lambda_H$, EWSB will
occur at weak coupling. Electroweak observables in this model, such as the
oblique parameters $S$ and $T$ or the Higgs spectrum, are thus calculable.

In \cite{fathiggs} the Higgs spectrum was calculated for a range of SUSY
breaking parameters. The lightest Higgs was indeed found to be heavier than
conventional MSSM values, easily reaching 350 GeV or higher. A distinct
signal for the Higgs spectrum in our models is that the pseudo-scalar Higgs
is always heavier than the charged one whereas in the MSSM the converse is
alway true. The $S$ and $T$ parameters were also calculated and were found to
be within the 1$\sigma$ allowed region for a wide range of parameters.

EWSB may be communicated to the matter sector by a scalar version of an ETC
sector. We add heavy fundamental Higgses, $\varphi_{u,d}$ and
$\bar\varphi_{u,d}$, that couple both to the $T$'s and to the MSSM
\begin{eqnarray}
  W_{f} &=& M_{f} (\varphi_u \bar{\varphi}_u + \bar{\varphi}_d
  \varphi_d) + \bar{\varphi}_d (T T^4)
  + \bar{\varphi}_u (T T^3) \nonumber \\
  &&+ h_u^{ij} Q_i u_j \varphi_u + h_d^{ij} Q_i d_j \varphi_d
  + h_e^{ij} L_i e_j \varphi_d.
\end{eqnarray}
At the scale $M_f$, the heavy Higgses can be integrated out, leaving a direct
coupling between the MSSM and the right combination of $T$'s that becomes the
composite Higgs once the theory confines. The effective Yukawa to the fat
Higgses is $\frac{h_{u,d}}{4\pi}\frac{\Lambda_H}{M_f}$. If
$M_f\sim\Lambda_H$ this may present a problem in generating a large top
mass.

In order to avoid fine tuning the SUSY breaking scale in this theory must be
of order $\lambda v_0$ which is set by supersymmetric scales. This problem is
reminiscent of the $\mu$ problem of the MSSM.  A more complete model was
introduced in~\cite{fathiggs} in order to relate the SUSY breaking scale to
$m$, $\Lambda_H$ and also $M_f$. This was done by adding a fourth flavor of
$T$'s which results in near conformal behavior.  The new walking dynamics is
also beneficial to enhancing the effective MSSM Yukawa couplings by anomalous
dimensions as in walking technicolor. However, in this case the anomalous
dimensions are calculable due to their relation to anomaly free $R$-charges. 

Finally, lets revisit the possibility of unification in this model. Below the
scale $\Lambda_H$ the matter content of our model is simply that of the
NMSSM. The running of the 3-2-1 gauge couplings follow their regular
trajectories. At $\Lambda_H$ we must match the low energy theory to the UV
completion. Due to holomorphy, the matching equation between the low and high
energy theories is set by bare parameters~\cite{HitoshiNima}. This leads us
to believe that the threshold corrections at $\Lambda_H$ are small. Above
$\Lambda_H$ the $T$'s must substitute the composite Higgses in the running.
It is amusing that the $T$'s contribute to the one-loop beta functions of the
MSSM exactly as two Higgs doublets do, which leaves us on the MSSM
trajectory.  However, we have also added the fundamental Higgses
$\varphi_{u,d}, \bar\varphi_{u,d}$, as well as new fields that take part in
the $W_{decouple}$,  all of which contribute as three more pairs of Higgs
doublets. This deviation from the MSSM running may be corrected. For example,
adding three pairs of heavy triplets with $Y=\pm 1/3$, completes the
``$SU(5)$ multiplets'' and brings us back on track. Though this remedy may
seem somewhat contrived\footnote{For a possibly more attractive solution with
a slimmer Higgs see~\cite{newfat}.}, we note that the possibility of even
talking about unification \emph{above} a compositeness scale may be viewed as
significant progress.

\small
\section{Acknowledgements}
I would like to thank my collaborators- Graham Kribs, Daniel Larson and
Hitoshi Murayama. Many thanks to Zacharia Chacko who suggested I go to SUSY
and for pointing out how many 10 minute intervals there were in February.
This work was supported in part by the DOE under contracts DE-FG02-90ER40542
and DE-AC03-76SF00098 and in part by NSF grant PHY-0098840.

\small
\bibliographystyle{plain}

\end{document}